\documentclass{aa} 
\usepackage{natbib}
\usepackage{graphicx} 
\usepackage{txfonts}
\usepackage{longtable} 
\usepackage{supertabular} 
\begin{document}

\newcommand{\sv}{$\sigma_{los}$} 
\newcommand{\rv}{$r_{200}$} 
\newcommand{\vv}{${\rm v}_{200}$} 
\newcommand{\mv}{$M_{200}$} 
\newcommand{\ks}{km~s$^{-1}$} 

\title{The orbital velocity anisotropy of cluster galaxies:
  evolution} \author{A. Biviano\inst{1} \and B.M. Poggianti\inst{2}}

\offprints{biviano@oats.inaf.it}

\institute{INAF/Osservatorio Astronomico di Trieste, via G. B. Tiepolo 11, I-34131, Trieste, Italy \and 
INAF/Osservatorio Astronomico di Padova, vicolo Osservatorio 5, I-35122, Padova, Italy}

\date{Received / Accepted}

\abstract{In nearby clusters early-type galaxies follow isotropic
  orbits. The orbits of late-type galaxies are instead characterized
  by slightly radial anisotropy.  Little is known about the orbits of
  the different populations of cluster galaxies at redshift $z >
  0.3$.}{We investigate the redshift evolution of the orbits of
  cluster galaxies.}{We use two samples of galaxy clusters spanning
  similar (evolutionary corrected) mass ranges at different
  redshifts. The sample of low-redshift ($z \sim 0.0-0.1$) clusters is
  extracted from the ESO Nearby Abell Cluster Survey (ENACS)
  catalog. The sample of high-redshift ($z \sim 0.4-0.8$) clusters is
  mostly made of clusters from the ESO Distant Cluster Survey
  (EDisCS). For each of these samples, we solve the Jeans equation for
  hydrostatic equilibrium separately for two cluster galaxy
  populations, characterized by the presence and, respectively,
  absence of emission-lines in their spectra ('ELGs' and 'nELGs'
  hereafter).  Using two tracers of the gravitational potential allows
  to partially break the mass--anisotropy degeneracy which plagues
  these kinds of analyses.}{We confirm earlier results for the nearby
  cluster sample. The mass profile is well fitted by a Navarro, Frenk
  \& White (NFW) profile with concentration $c=4$.  The mass profile
  of the distant cluster sample is also well fitted by a NFW profile,
  but with a slightly lower concentration, as predicted by
  cosmological simulations of cluster-sized halos. While the mass
  density profile becomes less concentrated with redshift, the number
  density profile of nELGs becomes more concentrated with redshift. In
  nearby clusters, the velocity anisotropy profile of nELGs is close
  to isotropic, while that of ELGs is increasingly radial with
  clustercentric radius. In distant clusters the projected phase-space
  distributions of both nELGs and ELGs are best-fitted by models with
  radial velocity anisotropy.}{No significant evolution is detected
  for the orbits of ELGs, while the orbits of nELGs evolve from radial
  to isotropic with time.  We speculate that this evolution may be
  driven by the secular mass growth of galaxy clusters during their
  fast accretion phase. Cluster mass density profiles and their
  evolution with redshift are consistent with predictions for
  cluster-sized halos in $\Lambda$ Cold Dark Matter cosmological
  simulations. The evolution of the nELG number density profile is
  opposite to that of the mass density profile, becoming less
  concentrated with time, probably a result of the transformation of
  ELGs into nELGs.}

\keywords{Galaxies: clusters: general -- Galaxies: kinematics and dynamics}

\titlerunning{The orbits of cluster galaxies: evolution}
\authorrunning{Biviano \& Poggianti}

\maketitle

\section{Introduction}
\label{s:intro}
The distributions of cluster early-type and late-type galaxies (ETGs
and LTGs hereafter) have long been known to be different
(\citealt{Dressler80}; see also \citealt{Biviano00} for a review).
Most striking, and hence the first to have been discovered, is the
difference in the {\em spatial} distributions, ETGs living in higher
density regions than LTGs, the so-called ``morphology-density
relation'' (MDR hereafter). In relaxed clusters density is an almost
monotonic decreasing function of clustercentric radius, hence the
relation is often described as a morphology-radius relation
\citep[e.g.][]{WGJ93}. The MDR has been found to exist in rich
clusters up to redshift $z \sim 1$ \citep{Postman+05,Smith+05},
  but the global fractions of ETGs and LTGs evolve with $z$. The
fraction of ETGs in clusters decreases with $z$ quite rapidly up to $z
\sim 0.5$ \citep{Dressler+97,Fasano+00,Postman+05,Smith+05}, then
seems to flatten out to $z \sim 1$ \citep{Desai+07}.

Another aspect of the segregation of different galaxy populations in
clusters is the difference in the velocity distributions of ETGs and
LTGs. In clusters, the velocity distribution of LTGs is broader than
that of ETGs, i.e. LTGs have a larger line-of-sight (los hereafter)
velocity dispersion (\sv~ hereafter) than ETGs
\citep{Tammann72,MD77,Sodre+89,Biviano+92}. The difference is not only
in the {\em global} \sv, but also in the \sv-profiles, 'hotter' and
steeper (at least in the inner regions) for the LTGs
\citep{Carlberg+97-equil,Biviano+97,ABM98}.

While the $z$-evolution of the MDR has been extensively studied and
described, less is known about the $z$-evolution of the morphological
segregation in velocity space, because of the difficulty of obtaining
large spectroscopic samples of cluster galaxies at high-$z$. As
remarked by \citet{Biviano06}, the \sv-profiles of both ETGs
and LTGs in $z \sim 0.07$ clusters \citep[from ENACS, the ESO Nearby Abell
Cluster Survey,][]{Katgert+96,Biviano+97} are remarkably similar to the
profiles of red, and respectively, blue galaxies in $z \sim 0.3$
clusters \citep[from CNOC, the Canadian Network for Observational
Cosmology survey,][]{Carlberg+97-equil}, but little is known about the
evolution at still higher $z$. 

Instead of looking separately at the spatial and velocity segregation
of cluster galaxies, it is possible to use the joint information
coming from their 2-dimensional projected phase-space distribution in
the los velocity vs. clustercentric radius diagram. The projected
phase-space distribution of a given class of cluster galaxies is the
observable that enters the Jeans equation for the equilibrium of a
galaxy system \citep[see, e.g.][]{BT87,Biviano08}, hence separating
different cluster galaxy populations on the basis of their projected
phase-space distributions is a way of identifying different,
independent tracers of the cluster gravitational potential.
\citet{Biviano+02} have shown that cluster ETGs and LTGs have
significantly different projected phase-space distributions;
\citet{KBM04} have then used cluster ETGs to determine the average
mass profile of massive clusters.  In order to do that, they first had
to constrain the velocity anisotropy profile of ETGs. They did so by
comparing the velocity distribution of ETGs with distribution function
models from \citet{vanderMarel+00}, and found that ETGs move on nearly
isotropic orbits. On the other hand, LTGs were found to move on
slightly radially anisotropic orbits, with an increasing radial
anisotropy at larger clustercentric radii \citep{BK04}.

At higher redshift, in the CNOC clusters, red galaxies were found to
move along nearly isotropic orbits \citep{vanderMarel+00}, similarly
to ETGs in nearby ENACS clusters, and blue galaxies were found to be
in equilibrium within the cluster potential, despite having a
different projected phase-space distribution from that of red galaxies
\citep{Carlberg+97-equil}. Although a solution for the velocity
anisotropy of the blue CNOC cluster galaxies has not been derived, the
similarity of their projected phase-space distribution to that of LTGs
in ENACS suggests that they similarly move on slightly radial orbits
\citep{Biviano06,Biviano08}.

Hence, while there is significant evolution in the relative fractions
of early-type (red) and late-type (blue) cluster galaxies, their
orbital anisotropies do not seem to evolve over the same 0--0.3
redshift range.  \citet{Benatov+06} do claim significant orbital
evolution for the {\em whole} cluster galaxy population, on the basis
of the analyses of three low-$z$ and two $z \sim 0.2-0.3$ clusters.
Specifically, they find galaxies in the higher-$z$ clusters to have
more radially anisotropic orbits than galaxies in the lower-$z$
clusters. Since they consider all cluster galaxies together in their
analysis, and since LTGs are known to be characterized by radial
orbital anisotropy, \citet{Benatov+06}'s result could be explained by
the increasing fraction of LTGs with $z$, without the need for any
evolution of the orbits of either ETGs or LTGs.

The lack of a significant evolution in the orbital anisotropy of,
separately, early-type (red) and late-type (blue) cluster galaxies
from $z \sim 0.05$ to $z \sim 0.3$, coupled with the significant
evolution in the relative fractions of these two populations over the
same redshift range is intriguing. It suggests that the change of
class from blue, LTG, to red, ETG, goes together with the orbital
change, from moderately radial to isotropic. The mechanisms by which
this evolution occurs are not known. To gain further insight into this
evolutionary process, in this paper we extend the analysis of galaxy
orbits in clusters to higher redshifts than examined so far. We base
our analysis on the sample of clusters from the ESO Distant Cluster
Survey \citep[EDisCS,][]{White+05}, that span the redshift range
$\simeq 0.4$--1.0.

We adopt $H_0=70$ km~s$^{-1}$~Mpc$^{-1}$, $\Omega_m=0.3$,
$\Omega_{\Lambda}=0.7$ throughout this paper.

\section{Two data-sets} 
\label{s:data}
The high-$z$ (distant) cluster galaxies data-set used in this paper
has been gathered in the EDisCS \citep{White+05,Poggianti+06}, a
survey of 20 fields containing galaxy clusters in the $z$-range
0.4--1.0.  Multi-band optical and near-infrared photometry for these
fields has been obtained using VLT/FORS2 \citep{White+05}, and
NTT/SOFI (Arag{\'o}n-Salamanca et al., in prep.). Imaging with the ESO
WFI has also been obtained as well as HST/ACS mosaic images for 10
cluster fields \citep{Desai+07} and deep optical spectroscopy for 18
cluster fields using VLT/FORS2 \citep{Halliday+04,MilvangJensen+08}.

Three of the 18 EDisCS clusters \citep[Cl~1103, Cl~1119, and Cl~1420,
  see Table~1 in][]{Poggianti+06} have very low-masses with velocity
dispersion \sv $<250$ \ks, more typical of groups than clusters. We
exclude them from our data-set since including them would break the
expected homology of cluster mass profiles that is a requirement for
the stacking procedure (see Sect.~\ref{s:stacked}).  We are thus left
with 15 EDisCS clusters.  In order to increase our data-set we add
four clusters from the MORPHS data-set
\citep{Dressler+99,Poggianti+99} with masses in the same range covered
by the 15 EDisCS clusters. These clusters \citep[Cl~0024, Cl~0303,
  Cl~0939, and Cl~1601; see Table~2 in][]{Poggianti+06} have
sufficiently wide spatial coverage ($>0.5 \, r_{200}$)\footnote{The
  virial radius \rv~ is the radius within which the enclosed average
  mass density of a cluster is 200 times the critical density.

  The virial mass \mv~ is the mass enclosed within a sphere of radius
  \rv.

The circular velocity is defined from the previous two quantities as
\vv=$(G$ \mv$/$\rv$)^{1/2}$.}, needed for the determination of the
mass profile, and homogeneous photometry, needed for the determination
of the radial incompleteness (see Sect.~\ref{s:stacked}).

As a reference low-$z$ (nearby) sample of clusters, we use the 59
ENACS clusters studied in detail by \citet{Biviano+02,BK04,KBM04}. A
full description of the ENACS can be found in
\citet{Katgert+96,Katgert+98}.

We identify and reject interlopers in both the high- and the low-$z$
clusters following the procedure described in \citet{Biviano+06}, which
is based on the identification of significant gaps in redshift space
\citep{Girardi+93} and on further removal of unbound galaxies
identified in projected phase-space \citep{dHK96}, a procedure
validated via the comparison with clusters extracted from numerical
simulations \citep{Biviano+06,Wojtak+07}. On the remaining cluster
members we determine cluster los velocity dispersions \sv~ using the
robust biweight estimator \citep{BFG90}. Finally, we determine \mv~
cluster masses adopting the scaling \sv--\mv~ relation of
\citet[][app. A]{MM07}, which is in good agreement with the phenomenological
relation derived by \citet{Biviano+06} for a sample of cluster-sized
halos extracted from cosmological simulations.

In summary, our data-set consists of 19 distant clusters from $z=0.393$
to $0.794$ with a mean (median) redshift $z=0.56 (0.54)$, and 59
nearby clusters from $z=0.035$ to $0.098$ with a mean (median)
redshift $z=0.07 (0.06)$. The distant clusters span the \mv~
mass-range $0.7-13.6 \times 10^{14} \, M_{\odot}$, with a
mean (median) mass of $2.8 (4.4) \times 10^{14} \, M_{\odot}$. The nearby
clusters span the \mv~ mass-range $0.4-20.5 \times 10^{14} \,
M_{\odot}$, with a mean (median) mass of $5.9 (5.7) \times 10^{14} \,
M_{\odot}$. Assuming a cluster mass accretion rate of $\sim 0.08 \,
M_{200}(z=0)/$Gyr \citep{Adami+05}, the average mass of our
high-$z$ cluster sample is expected to increase to about the average
mass of our low-$z$ cluster sample in the time that separates the two
cosmic epochs ($\sim 5$ Gyr). A similar,
albeit somewhat larger, evolution in mass is also predicted
theoretically \citep{LC09}. In this sense, we are comparing similar
objects observed at two different cosmic epochs.

The low-$z$ and high-$z$ cluster samples are presented in
Table~\ref{t:lowz} and \ref{t:highz}, respectively. In col.(1) we list
the cluster name \citep[following the short-name convention of][for
  the high-$z$ clusters]{Poggianti+06}, in col.(2) its mean redshift,
in col.(3) the cluster \mv~ in $10^{14} M_{\odot}$ units, in col.(4)
the number of galaxies without emission lines in their spectra (nELGs,
hereafter) in the radial range $0.05 \leq R/r_{200} \leq 1$, in
col.(5) the number of galaxies with emission lines (ELGs, hereafter)
in the same radial range, and in col.(6) the radial distance from the
cluster center of the most distant galaxy in the sample, in units of
$r_{200}$. The ELG classification for the low-$z$ cluster sample is
described in \citet{Katgert+96}. For the high-$z$ cluster sample we
classify ELGs the EDisCS galaxies with an [OII] equivalent width $\geq
3$~\AA~ or with any other line in emission \citep{Poggianti+06} and the
MORPHS galaxies with a spectral type different from 'k', 'k+a', and
'a+k' \citep{Poggianti+99}.

\begin{table}
\centering
\caption[]{The low-$z$ cluster sample.}
\label{t:lowz}
\begin{tabular}{lrrrrr}
\hline
Id. & $\overline{z_c}$ & $M_{200}$ & $N_{nELG}$ & $N_{ELG}$ & $R_{max}/r_{200}$ \\
\hline
A0013  & 0.0932 &  9.70 &  33 &   2 &  0.98 \\
A0087  & 0.0538 &  9.14 &  25 &   2 &  0.46 \\
A0119  & 0.0433 &  5.13 &  94 &   5 &  0.97 \\
A0151a & 0.0402 &  0.92 &  18 &   4 &  1.00 \\
A0151b & 0.0523 &  5.70 &  30 &   7 &  0.99 \\
A0151c & 0.0982 &  6.99 &  24 &   3 &  0.99 \\
A0168  & 0.0440 &  1.93 &  58 &   4 &  0.98 \\
A0295  & 0.0416 &  0.38 &  24 &   1 &  0.81 \\
A0514  & 0.0712 &  9.12 &  67 &  11 &  0.94 \\
A0524  & 0.0775 &  6.98 &  12 &  12 &  0.73 \\
A0548a & 0.0413 &  4.92 &  71 &  33 &  0.99 \\
A0548b & 0.0421 &  7.71 &  89 &  21 &  1.00 \\
A0754  & 0.0558 & 14.05 &  35 &   0 &  0.49 \\
A0957  & 0.0455 &  3.79 &  31 &   0 &  0.47 \\
A0978  & 0.0555 &  1.71 &  30 &   3 &  0.98 \\
A1069  & 0.0664 & 11.21 &  32 &   0 &  0.52 \\
A1809  & 0.0806 &  6.29 &  29 &   0 &  0.80 \\
A2040  & 0.0466 &  4.23 &  32 &   3 &  0.57 \\
A2048  & 0.0977 &  3.94 &  22 &   1 &  0.87 \\
A2052  & 0.0355 &  5.15 &  28 &   2 &  0.41 \\
A2361  & 0.0597 &  0.52 &  11 &   6 &  0.99 \\
A2401  & 0.0561 &  1.48 &  20 &   1 &  0.80 \\
A2569  & 0.0797 &  1.55 &  27 &   2 &  0.99 \\
A2734  & 0.0607 &  2.68 &  59 &   1 &  1.00 \\
A2799  & 0.0624 &  1.06 &  25 &   4 &  0.99 \\
A2800  & 0.0626 &  0.89 &  21 &   6 &  0.98 \\
A2819  & 0.0743 &  0.96 &  30 &   2 &  0.84 \\
A2819  & 0.0862 &  0.61 &  17 &   4 &  0.96 \\
A2911  & 0.0800 &  0.92 &  18 &   0 &  0.88 \\
A3093  & 0.0826 &  0.94 &  10 &   2 &  0.70 \\
A3094  & 0.0668 &  3.84 &  41 &  14 &  0.98 \\
A3111  & 0.0773 &  6.20 &  32 &   3 &  0.85 \\
A3112  & 0.0747 & 11.78 &  51 &  15 &  0.96 \\
A3122  & 0.0639 &  6.53 &  59 &  13 &  0.99 \\
A3128  & 0.0598 &  6.10 &  95 &  17 &  1.00 \\
A3151  & 0.0678 &  5.84 &  30 &   2 &  0.69 \\
A3158  & 0.0590 & 13.89 &  93 &   8 &  0.97 \\
A3194  & 0.0970 &  6.82 &  22 &   7 &  0.98 \\
A3202  & 0.0691 &  1.15 &  21 &   3 &  0.98 \\
A3223  & 0.0599 &  2.93 &  50 &   2 &  0.99 \\
A3341  & 0.0379 &  2.46 &  48 &  11 &  0.99 \\
A3354  & 0.0586 &  0.70 &  22 &   3 &  1.00 \\
A3365  & 0.0929 & 20.47 &  27 &   5 &  0.55 \\
A3528  & 0.0546 & 12.49 &  26 &   0 &  0.78 \\
A3558  & 0.0486 & 15.15 &  60 &   9 &  0.98 \\
A3559  & 0.0477 &  1.07 &  21 &   4 &  0.99 \\
A3562  & 0.0488 & 10.07 &  55 &  12 &  1.00 \\
%
A3651  & 0.0595 &  3.99 &  51 &   3 &  0.99 \\
A3667  & 0.0554 & 15.22 &  83 &   9 &  0.99 \\
A3691  & 0.0867 &  4.65 &  28 &   1 &  0.88 \\
A3705  & 0.0890 & 15.95 &  24 &   3 &  0.62 \\
A3764  & 0.0748 &  2.71 &  23 &  10 &  1.00 \\
A3806  & 0.0761 &  7.17 &  50 &   9 &  1.00 \\
A3822  & 0.0754 & 12.41 &  60 &  13 &  1.00 \\
A3825  & 0.0746 &  4.67 &  46 &   4 &  0.98 \\
A3827  & 0.0978 & 19.43 &  19 &   1 &  0.89 \\
A3879  & 0.0666 &  1.22 &  26 &   8 &  0.92 \\
A3921  & 0.0930 &  1.66 &  20 &   3 &  0.98 \\
A4010  & 0.0948 &  3.27 &  21 &   6 &  0.92 \\
\hline
\end{tabular}
\end{table}

\begin{table}
\centering
\caption[]{The high-$z$ cluster sample.}
\label{t:highz}
\begin{tabular}{lrrrrr}
\hline
Id. & $\overline{z_c}$ & $M_{200}$ & $N_{nELG}$ & $N_{ELG}$ & $R_{max}/r_{200}$ \\
\hline
Cl 1018 & 0.4734 &  1.15 &   9 &  11 &  0.94 \\
Cl 1037 & 0.5782 &  0.74 &   2 &  12 &  0.93 \\
Cl 1040 & 0.7044 &  0.73 &   3 &  11 &  0.95 \\
Cl 1054-11 & 0.6976 &  1.56 &   8 &  16 &  0.92 \\
Cl 1054-12 & 0.7500 &  0.91 &  10 &   7 &  0.95 \\
Cl 1059 & 0.4563 &  1.35 &  14 &  15 &  0.82 \\
Cl 1138 & 0.4797 &  3.61 &   8 &  33 &  0.82 \\
Cl 1202 & 0.4239 &  1.75 &  11 &   4 &  0.86 \\
Cl 1216 & 0.7940 &  9.97 &  31 &  27 &  0.89 \\
Cl 1227 & 0.6360 &  1.83 &   3 &   9 &  0.95 \\
Cl 1232 & 0.5420 & 12.34 &  34 &  13 &  0.58 \\
Cl 1301 & 0.4832 &  3.27 &  10 &  20 &  0.96 \\
Cl 1353 & 0.5877 &  1.94 &   7 &   6 &  0.98 \\
Cl 1354 & 0.7618 &  2.84 &   2 &   8 &  0.99 \\
Cl 1411 & 0.5200 &  3.45 &  12 &   6 &  0.92 \\
Cl 0024 & 0.3933 & 11.67 &  42 &  32 &  0.56 \\
Cl 0303 & 0.4185 &  6.96 &  19 &  18 &  0.64 \\
Cl 0939 & 0.4076 & 13.55 &  28 &   9 &  0.51 \\
Cl 1601 & 0.5401 &  3.76 &  40 &   6 &  0.79 \\
\hline
\end{tabular}
\end{table}

\section{The construction of the stacked cluster samples}
\label{s:stacked}
In order to be able to analyse the cluster mass and velocity
anisotropy profiles, the available spectroscopic data of individual
clusters from our data-sets are not sufficient. We need to stack all
the clusters from each of the two samples together. Stacked cluster
samples have been used successfully in several analyses of the
properties of clusters
\citep[e.g.][]{MD77,Biviano+92,Carlberg+97-mprof,vanderMarel+00,KBM04,Rines+03}.
The validity of this approach is supported by the results of
cosmological numerical simulations that predict cosmological halos to
be characterized by the same, universal mass density profile
\citep{NFW97}. Even if the mass density profiles of cosmological halos
do depend on their mass \citep[see, e.g.,][]{NFW97,Dolag+04}, this
dependence is very mild, and the profiles are very similar for halos
with masses within about two decades around the average cluster-like
halo mass.  In order to ensure homology of our distant cluster
data-set we rejected three very low-mass clusters from the initial
sample (see Sect.~\ref{s:data}).

The observables on which the analysis is based (see
  Sect.~\ref{s:method}) are the galaxy projected clustercentric
distances, $R$, and the galaxy los velocities in their cluster rest
frames \citep{HN79}, ${\rm v}_{rf} \equiv ({\rm v}-\overline{{\rm
    v}_c})/(1+\overline{z_c})$, where ${\rm v}$ are the observed
galaxy velocities, and $\overline{z_c}, \overline{{\rm v}_c}$ the
average cluster redshift and los velocity, respectively.  For the
cluster centers we use the position of the X-ray surface brightness
peak, or, if this is unavailable, the position of the cluster
brightest cluster galaxy (BCG hereafter). Since different
determinations of a cluster center typically differ by less than 100
kpc \citep{Adami+98-7}, precise centering is not very important for
our dynamical analysis \citep[see also][]{Biviano+06}.

In order to stack clusters together it is necessary to adopt
appropriate cluster-dependent scalings for the $R$ and ${\rm v}_{rf}$
quantities. Stacking the clusters in physical units would have the
unwanted consequence of mixing up the virialized regions of some
clusters with the unvirialized, external regions of others. Radii and
velocities are therefore scaled by the clusters virial radii, \rv, and
circular velocities, \vv, so that the dynamical analysis on the
stacked cluster is done in the normalized units $R_n \equiv R/r_{200}$
and ${\rm v}_n \equiv {\rm v}_{rf}/{\rm v}_{200}$.

In our analysis we consider only the virialized cluster regions, i.e.
galaxies with $R_n \leq 1$. The Jeans method \citep{BT87} that we
adopt here to determine cluster mass and anisotropy profiles is in
fact not valid outside the region where dynamical equilibrium is
likely to hold. Moreover, we also exclude the very central cluster
regions ($R_n < 0.05$) in order to account for the positional
  uncertainties of cluster centers, and in order not to include the
centrally located BCGs in our sample. BCGs are probably built up via
merger processes even quite recently \citep[see,
  e.g.,][]{DLB07,Ramella+07,RFV07}.  Including these galaxies would
therefore probably invalidate the Jeans analysis in which dissipation
processes are not included \citep{MF96}.

In the selected radial range $0.05 \leq R_n \leq 1$ there are 556
galaxies in the distant stacked cluster, and 2566 in the nearby one.
We further split these samples in two by considering nELGs and
ELGs separately.  There are 293 nELGs and 263 ELGs in the
distant stacked cluster, and 2226 nELGs and 340 ELGs in the nearby
one. The much larger fraction of ELGs in the distant stacked cluster
compared to the nearby one is a known feature of the evolution of the
cluster galaxy population fractions
\citep{Dressler+99,Poggianti+99,Poggianti+06}.

Neither the low-$z$ nor the high-$z$ cluster samples are
spectroscopically complete to a given magnitude. Incompleteness is not
a problem in the dynamical analysis as far as it is the same at all
radii. In fact, the normalization of the galaxy number density profile
cancels out in the Jeans procedure (see eqs.(\ref{eq:sigmar}) and
(\ref{eq:sigmalos}) in Sect.~\ref{s:method}) and the shape of the galaxy
number-density and velocity-dispersion profiles are known to be
independent from the galaxy luminosities, at least for absolute
magnitudes $M_R > -22.8$ \citep{Biviano+02}. Galaxies brighter than
$M_R=-22.8$ are mostly BCGs, and have been mostly excluded from our
samples by removing the very central cluster regions, $R_n < 0.05$.
While there are several indications that dwarf galaxies in clusters
have a different phase-space distribution from bright galaxies
\citep[e.g.,][]{Lobo+97,Kambas+00,OSR02,Popesso+06-4}, this is
irrelevant here since dwarf galaxies are not present in our samples.

If incompleteness does depend on radius, a correction must be applied.
In fact, a radially-dependent incompleteness would change not only the
normalization of the galaxy density profile but also its shape.  For
the ENACS, from which our local sample is extracted, it has been shown
that the spectroscopic incompleteness is not radial dependent
\citep{Katgert+98}. On the other hand, for the high-$z$ cluster sample
spectroscopic incompleteness {\em is} a function of radius, although
not a strong one, and must be corrected for \citep{Poggianti+06}.  The
incompleteness correction for each cluster in our high-$z$ sample is
done by applying the geometrical weights with the method described in
\citet[][Appendix A]{Poggianti+06}.

Another kind of radial incompleteness results from the stacking
procedure when the clusters that enter the stacked sample do not cover
the same radial range, i.e. they have not all been sampled out to the
same limiting aperture.  This happens for both our local and distant
samples (the apertures are listed in the last column of
Tables~\ref{t:lowz} and \ref{t:highz}).  A radical way of addressing
this problem would be to stack clusters only out to the minimum among
all clusters apertures.  In order to make the most efficient use of
our sample, and explore the clusters dynamics out to \rv, we prefer to
follow another approach. We apply a correction for the fact that not
all clusters contribute at all radii. This correction is based on the
relative individual cluster contributions to the total number of
galaxies in the stacked sample, estimated at the radii where the
clusters are still sampled. E.g. if at a given radius $R_{n,m}$, there
are $m$ clusters that have not been sampled out to that radius, their
contributions at radii $R_n>R_{n,m}$ is accounted for by applying a
correction factor $1/(1-\sum_{i=1}^{m} f_i)$ to the total galaxy
counts in the stacked cluster, where $f_i$ is the fraction of galaxies
contributed to the total sample by cluster $i$ in the radial range
where the cluster is sampled.  The method is similar to the one
described in \citet{MK89} and has been applied to the ENACS sample in
\citet{Biviano+02} and \citet{KBM04}.

\begin{figure}
\centering \resizebox{\hsize}{!}{\includegraphics{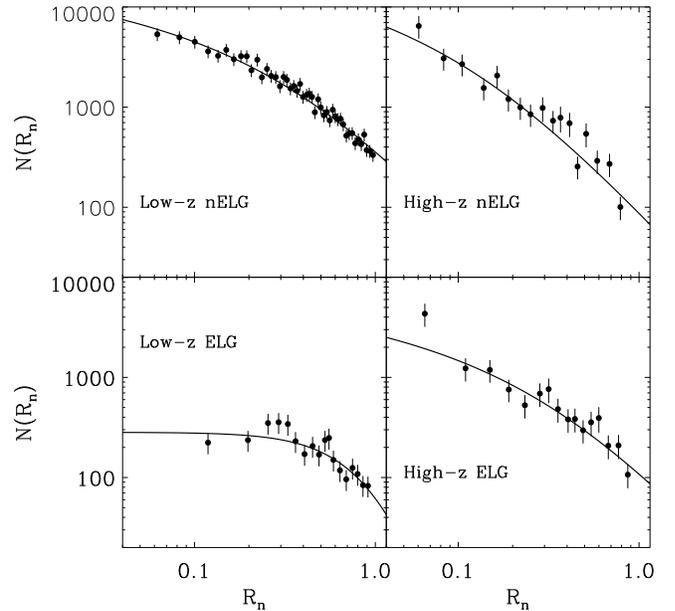}}
\caption{The projected number density profiles, $N(R_n)$, of the nELGs
  and ELGs (upper and lower panel, respectively) for the low- and
  high-$z$ stacked clusters (left and right panel, respectively). Solid
  lines represent best-fit models to the data, i.e. projected NFW
  models for nELGs and high-$z$ ELGs, and the core model for low-$z$
  ELGs (see Sect.~\ref{s:enacs} and \ref{s:ediscs}).}
\label{f:nprofs}
\end{figure}

\begin{figure}
\centering \resizebox{\hsize}{!}{\includegraphics{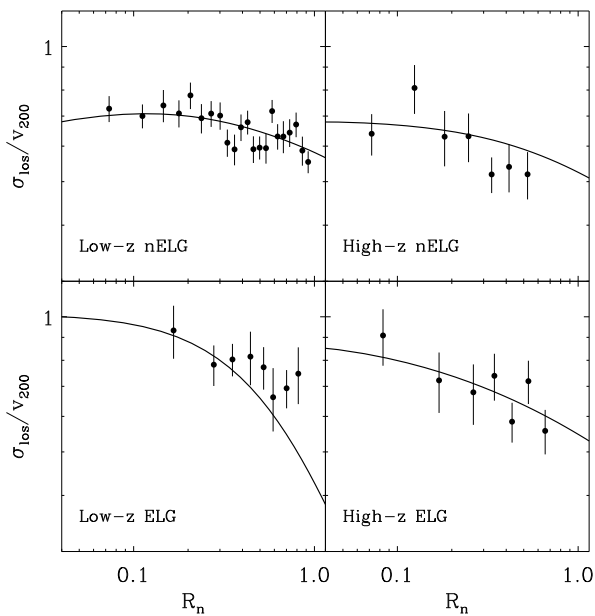}}
\caption{The los velocity dispersion profiles \sv$(R_n)$, of
the nELGs and ELGs (upper and lower panel, respectively) for
the low- and high-$z$ stacked cluster (left and right panel,
respectively). Solid lines represent best-fit models to the
data, i.e. $c=4.0$ NFW mass model, $a=3.6$ (respectively $a=1.2$) 
OM velocity-anisotropy model
for low-$z$ nELGs (respectively, ELGs), and 
$c=3.2$ NFW mass model, $a=0.01$ M{\L} 
velocity-anisotropy model for high-$z$ nELGs
and ELGs (see Sect.~\ref{s:enacs} and \ref{s:ediscs}).}
\label{f:vprofs}
\end{figure}

In Fig.~\ref{f:nprofs} and \ref{f:vprofs} we display the projected
number density profiles, $N(R_n)$, and the los velocity dispersion
profiles, $\sigma_{los}(R_n)$, respectively, of nELGs and ELGs for our
local and distant stacked clusters.

\section{The dynamical analysis: method}
\label{s:method}
The method we adopt for the dynamical analysis of our two stacked
clusters is based on the standard spherically-symmetric Jeans analysis
\citep{BT87}. 

The assumption of spherical symmetry appears justified by the fact
that our samples combine many clusters together, irrespective of their
orientation \citep[see also][]{vanderMarel+00,KBM04}. The resulting
stacked clusters are spherically symmetric by construction, except if
the cluster selection process favors a preferential orientation. E.g.,
it has been argued that clusters selected because of the presence of
gravitational arcs are more likely to have their major axes aligned
along the los \citep[e.g.][]{Allen98}. Neither ENACS nor EDisCS
clusters were selected because of the presence of gravitational arcs,
and they are both likely to be representative of the overall cluster
populations in their mass and redshift ranges.  ENACS clusters are
drawn from the catalog of rich clusters in \citet{ACO89}. The mass
function derived using clusters from this catalog is similar to that
obtained using X-ray selected clusters \citep{Mazure+96}, hence rich
clusters from the \citet{ACO89} catalog appear to be representative of
all rich clusters in the nearby universe. EDisCS clusters were
optically selected from the highest surface brightness candidates in
the Las Campanas Distant Cluster Survey \citep{GZDN01}. From a
comparison with other published optical/X-ray cluster catalogs,
\citet{GZDN01} have shown that their detection method is able to
recover $>90$\% of the cluster population. Further support against a
possible orientation bias comes from the comparison of different
EDisCS cluster mass-estimates, obtained using the
distribution of cluster galaxies, gravitationally lensed images
\citep{Clowe+06,MilvangJensen+08} and the X-ray emission from the intra-cluster
plasma \citep{Johnson+06}.

In the spherically-symmetric Jeans analysis, our observables are the
galaxy number density profile $N(R_n)$ and the los velocity dispersion
profile $\sigma_{los}(R_n)$.  $N(R_n)$ is uniquely related to the
3-dimensional (3-d) galaxy number density profile $\nu(r_n)$ via the
Abel inversion equation,
\begin{equation}
\nu(r_n) = - \frac{1}{\pi} \int_{r_n}^{\infty} \frac{dN}{dR_n} 
\frac{dR_n}{\sqrt{R_n^2-r_n^2}},
\label{eq:abel}
\end{equation}
where $r_n \equiv r/r_{200}$ is the 3-d clustercentric radius in
normalized units.  The other observable, $\sigma_{los}(R_n)$ is
related to the cluster mass profile, $M(r_n)$, and the cluster
velocity anisotropy profile,
\begin{equation}
\beta(r_n) \equiv 1 - \frac{\rm{<} v_t^2 \rm{>}}{\rm{<} v_r^2 \rm{>}},
\label{eq:beta}
\end{equation}
where $\rm{<} v_t^2 \rm{>}$, $\rm{<} v_r^2 \rm{>}$ are the mean
squared tangential and radial velocity components, which reduce to
$\sigma_t^2$ and $\sigma_r^2$ respectively, in the absence of bulk
motions and net rotation. Given $M(r_n)$ and $\beta(r_n)$, the
observable $\sigma_{los}(R_n)$ follows through \citep{vanderMarel94}
\begin{equation}
\sigma_r^2(r_n) = G/\nu(r_n) \int_{r_n}^{\infty} 
\frac{\nu \, M}{\xi^2} \, \exp \left[ 2 \int_{r_n}^{\xi} \frac{\beta 
\, dx}{x}\right] d \xi,
\label{eq:sigmar}
\end{equation}
and \citep{BM82}
\begin{equation}
\sigma_{los}^2(R_n) = 2/N(R_n) \int_{R_n}^{\infty} 
\left( 1 - \beta \frac{R_n^2}{r_n^2}
\right) \frac{\nu \, \sigma_r^2 \, r_n \, dr_n}{\sqrt{r_n^2-R_n^2}},
\label{eq:sigmalos}
\end{equation}
where $G$ is the gravitational constant. Note that in practice the
upper limit of the integrals in the above equations is set to a finite
radius (typically $20$ in the integration units), large enough as to
ensure that the result of the integration does not change
significantly by pushing the limit to larger values.

It is therefore possible to adopt parameterized model representations
of $M(r_n)$ and $\beta(r_n)$ and determine the best-fit parameters by
comparing the observed $\sigma_{los}(R_n)$ profile with the predicted
one, using the $\chi^2$ statistics and the uncertainties on the
observed profile.

From the eqs. above it is however clear that different combinations of
the mass and anisotropy profiles can produce the same los velocity
dispersion profile, the so-called ``mass--anisotropy'' degeneracy.
Different methods exist to solve this degeneracy \citep[see,
  e.g.,][]{Merritt87,vanderMarel+00,LM03,WT06}. These methods are
effective for data-sets of $\sim 1000$ tracers of the gravitational
potential.  Given the smaller size of our distant clusters data-set we
here adopt another method, recently suggested by \citet{Battaglia+08}
for application to the case of dwarf galaxies.

The method of \citet{Battaglia+08} consists in considering not one,
but two different tracers of the cluster gravitational potential, so
that there are two observables (the los velocity dispersion profiles
of the two tracers) to solve for the two unknowns, $M(r_n)$ and
$\beta(r_n)$.  Since $M(r_n)$ must be the same for both tracers, but
$\beta(r_n)$ can in principle be different, the degeneracy is only
partially broken, however the constraints on the dynamics of the
system are significantly stronger than with a single tracer. Of
course, this method works only if the two tracers have different
projected phase-space distributions. This is the case of cluster nELGs
and ELGs, which we therefore adopt as our two populations of tracers.

In practice, we adopt parameterized models for $M(r_n)$ and
$\beta(r_n)$, solve eqs.(\ref{eq:abel}), (\ref{eq:sigmar}), and
(\ref{eq:sigmalos}) separately for nELGs and ELGs in each of the two
stacked clusters, and jointly compare the solutions to the observed
$\sigma_{los}(R_n)$ of the two galaxy populations. Say
$\sigma_{obs,i}$ and $\sigma_{model,i}$ the observed and predicted
$\sigma_{los}$ of a given population of tracers (nELG or ELG), in the
$i$-th of $m$ radial bins, and say $\delta_i$ the corresponding
1-$\sigma$ uncertainty on $\sigma_{obs,i}$. We find the best-fit
parameter of the model representing $M(r_n)$ and its uncertainties by
minimizing
\begin{equation}
\chi^2 = \chi^2_{nELG} + \chi^2_{ELG}
\label{eq:chi2sum}
\end{equation}
with 
\begin{equation}
\chi^2_{tracer} = \sum_{i}^{m} (\sigma_{obs,i} - \sigma_{model,i})^2/\delta_i^2.
\label{eq:chi2}
\end{equation}
Since $M(r_n)$ must be unique for the two populations, we then adopt
the joint best-fit solution for $M(r_n)$ and solve again
eqs.(\ref{eq:sigmar}) and (\ref{eq:sigmalos}) separately for nELGs and
ELGs to determine the best-fit parameters (and uncertainties) of their
$\beta$-profiles through the $\chi^2$ minimization, eq.(\ref{eq:chi2}),
i.e. we marginalize over $M(r_n)$ to constrain the two $\beta(r_n)$
solutions.

The choice of the models cannot be too generic nor too restrictive.
If it is too restrictive, we might find it difficult to find accurate
fits to the data. On the other hand, with small data-sets (such as
that of our high-$z$ sample) it would be difficult to obtain strong
constraints on models that are too generic or are characterized by too
many parameters. We let ourselves be guided in our choices by
theoretical expectations.

The distribution of mass within cosmological halos, such as galaxy
clusters, is a robust prediction of the Cold Dark Matter (CDM)
cosmological model. Numerical simulations have shown that all
cosmological halos are characterized by the same, universal, mass
density profile \citep{NFW97}, the so-called 'NFW' profile,
\begin{equation}
  \rho_{NFW} \propto (c r_n)^{-1} (1 + c r_n)^{-2},
\end{equation}
characterized by the concentration parameter $c$, a central cusp, and
an asymptotic slope of $-3$ at large radii.  While other
analytical forms have subsequently been proposed \citep[see,
  e.g.][]{Moore+99,Hayashi+04,Diemand+05}, the NFW profile provides an
acceptable fit to observed mass profiles of galaxy clusters both at
low and intermediate redshifts \citep[see,
  e.g.][]{vanderMarel+00,BG03,KBM04,Zappacosta+06}. It is therefore
rather straightforward to choose the NFW profile as our reference
model for $M(r)$.

We adopt the NFW profile also as a model for $\nu(r_n)$
\citep[actually, we fit $N(R_n)$ with the projected NFW profile,
  see][]{Bartelmann96,LM01}, but we do
{\em not} make the assumption that $\nu(r_n)$ and $M(r_n)$ are
characterized by the same NFW profile. I.e. we do not work in the
so-called light-traces-mass hypothesis. Whenever the NFW model does
not provide an acceptable fit to $\nu(r_n)$, we consider a different
model, with one additional parameter, to allow for a better fit. The model
we adopt in this case is the so-called $\beta$-model \citep{CF78},
\begin{equation}
N \propto [1+(R_n/R_c)^2]^{-\alpha},
\end{equation}
characterized by two parameters, the core-radius, $R_c$, and the
slope, $\alpha$. In order to avoid terminology confusion with the
velocity-anisotropy profile $\beta(r_n)$, in the following we refer to
the $\beta$-model as the 'core' model.

The model choice for the velocity-anisotropy profile $\beta(r_n)$ is less
straightforward. At variance with the case of the halo mass profile,
there is no claimed 'universal' velocity anisotropy profile. We
consider the two following models. One is the Mamon-{\L}okas
('M{\L}' hereafter) model \citep{ML05b}
\begin{equation}
\beta = 0.5 \, r_n/(r_n+a), 
\end{equation}
which has been
shown to provide a good fit to the velocity anisotropy profiles of
simulated cosmological halos. The other is the Osipkov-Merritt 
('OM' hereafter) model \citep{Osipkov79,Merritt85-df}
\begin{equation}
\beta = r_n^2/(r_n^2+a^2), 
\end{equation}
which provides a good fit to the observed velocity anisotropy profile
of late-type galaxies in the ENACS sample \citep{BK04}. Both the M{\L}
and the OM models depend on just one free parameter, the anisotropy
radius, $a$.

\section{Results}
\subsection{The nearby cluster sample}
\label{s:enacs}
The NFW profile (in projection) provides an acceptable fit to the nELG
$N(R_n)$, with $c=2.4$. This best-fit profile is shown in
the upper-left panel of Fig.~\ref{f:nprofs}. On the other hand, the
ELG $N(R)$ cannot be fitted by a (projected) NFW profile,
because the ELGs avoid the central cluster region. The wider spatial
distribution of the ELGs as compared to that of the nELGs clearly
reflects the well known MDR (see Sect.~\ref{s:intro}).  In order to
fit the rather flat ELG $N(R_n)$, we adopt the core model, and we find
an acceptable fit with $R_c=1.28$ and $\alpha=-3.2$ (see the
bottom-left panel of Fig.~\ref{f:nprofs}). The $N(R_n)$ best-fitting
models are Abel-inverted to provide the 3-d number density profiles
$\nu(r_n)$ that we use in the Jeans analysis.

\begin{figure}
\centering \resizebox{\hsize}{!}{\includegraphics{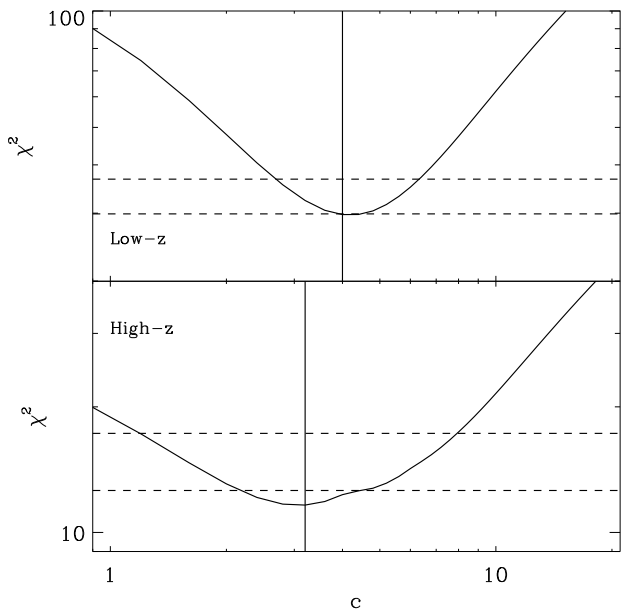}}
\caption{The $\chi^2$ values vs. the $c$ parameter of the NFW mass
  profile model for our low-$z$ (top panel) and high-$z$ (bottom
  panel) stacked clusters. The two dashed lines indicate the 68\% and
  90\% c.l., and the vertical lines indicate the best-fit $c$ values.}
\label{f:chi2c}
\end{figure}

The observed \sv-profiles of the nELGs and ELGs are shown in the
upper-left and, respectively, lower-left panels of Fig.~\ref{f:vprofs}.
The strikingly different behavior of the two \sv-profiles reflects the
well known morphological segregation in velocity space (see
Sect.~\ref{s:intro}). We then determine the best-fit concentration
parameter of a NFW $M(r_n)$ model by a joint $\chi^2$ comparison of
the observed nELG and ELG \sv-profiles with those predicted by the
Jeans analysis for the given mass model, and for either M{\L} or OM
$\beta(r_n)$ models. The best-fit solution is obtained using the OM
model for the velocity anisotropy profiles of nELGs and ELGs. The
best-fit value of the NFW concentration parameter is
$c=4.0_{(-0.1) -1.3}^{(+0.5) +2.3}$
(90\% confidence levels, c.l. in the following, 68\% c.l. in brackets).
The $\chi^2$ vs. $c$ solution is displayed in the top panel of
Fig.~\ref{f:chi2c}.

Marginalizing over this $M(r_n)$ best-fit solution we obtain the
best-fit $\beta(r_n)$ OM-model parameters $a=3.6_{-1.6}^{+13.4}$ (in
units of $r_{200}$) and $a=1.2_{-0.4}^{+1.2}$ for the nELG and ELG
populations, respectively. We only provide here 90\% c.l., since we
find no acceptable solutions at the 68\% c.l. for the ELGs. The
best-fit model \sv-profiles are shown as solid lines overlaid on the
observed, binned \sv$(R_n)$ in Fig.~\ref{f:vprofs} (left-hand panels).
Clearly, the fit to the ELG \sv$(R_n)$ is not excellent, but still
acceptable to within 90\% c.l.

\begin{figure}
\centering \resizebox{\hsize}{!}{\includegraphics{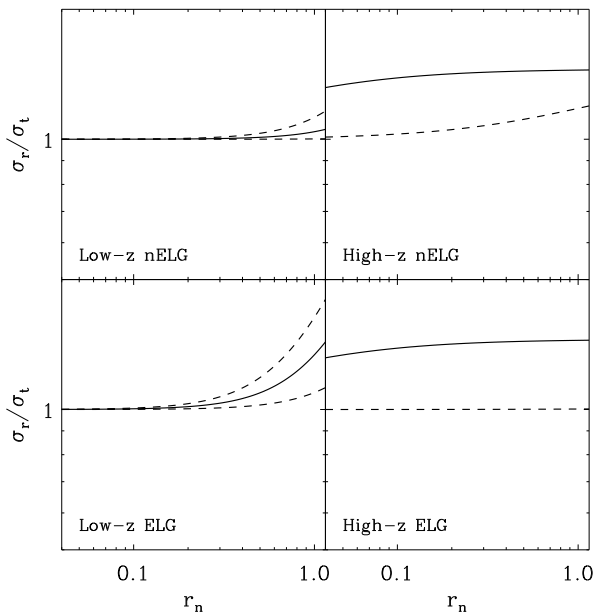}}
\caption{Best-fit velocity-anisotropy profile 
$\sigma_r/\sigma_t \equiv
    (1-\beta)^{-1/2}$ for nELGs (upper left
  panel) and ELGs (lower left panel) in low-$z$ clusters and for nELGs
  (upper right panel) and ELGs (lower right panel) in high-$z$
  clusters. The best-fit velocity-anisotropy model is OM for the
  low-$z$ sample and M{\L} for the high-$z$ sample.  Dashed lines
  indicate the 90\% c.l.
For the high-$z$ sample, the
  best-fit solutions are at the lower-limit of the interval considered
  in the $\chi^2$ minimization analysis, hence only the lower c.l. to
  $\beta(r_n)$ are shown.}
\label{f:beta}
\end{figure}

The solutions for the velocity-anisotropy profiles are displayed
in Fig.~\ref{f:beta} for the nELGs (upper-left panel) and the ELGs
(lower-left panel). Notice that the quantity displayed in
  Fig.~\ref{f:beta} is $\sigma_r/\sigma_t \equiv
    (1-\beta)^{-1/2}$.  The nELG velocity-anisotropy profile is
  consistent with fully isotropic orbits of this population of
  galaxies within the virial region, in agreement with the results
  obtained by \citet{KBM04}. On the other hand, the ELG orbits
    are approximately isotropic only out to $r_n \simeq 0.6$, and
  then become increasingly radial. The OM-model solution is very
  similar to the non-parametric velocity-anisotropy profile
  derived by \citet{BK04}. It is also quite similar to the
  velocity-anisotropy profiles of DM particles in cluster-size
  simulated halos at $z \approx 0$ \citep[e.g.][]{TBW97,Diaferio+01}.

For the sake of comparison with the distant cluster sample (see
Sect.~\ref{s:ediscs}) it is useful to also consider the solution
obtained for the M{\L} $\beta(r_n)$ model. With such a model, we
obtain an acceptable solution of the Jeans analysis for a NFW mass
profile with the same concentration value obtained using the OM
$\beta(r_n)$ model ($c=4$), although with a larger $\chi^2$
value. Marginalizing over the $c=4$ parameter, we then obtain the
best-fit $\beta(r_n)$ M{\L} model parameters $a=2.9_{-2.4}^{>+20}$ and
$a=1.5_{-0.8}^{+3.4}$ (90\% c.l.) for the nELG and ELG populations,
respectively. Similarly to what is obtained using the OM $\beta(r_n)$
model, also in this case the best-fit anisotropy radius of the nELG
population is about twice as large as that of the ELG population,
indicating that the ELG orbits in clusters are more radially elongated
than the nELG orbits.

\subsection{The distant cluster sample}
\label{s:ediscs}
Both the nELG and the ELG $N(R_n)$ are well fitted by projected NFW
profiles with $c=7.5$ and $c=2.7$, respectively.  These $N(R_n)$
best-fitting models are Abel-inverted to provide the 3-d number
density profiles $\nu(r_n)$ that we use in the Jeans analysis. The
binned profiles and their best-fit models are shown in the right-hand
panels of Fig.~\ref{f:nprofs}. Similarly to what is found in
the nearby cluster sample, ELGs have a wider spatial distribution than
nELGs, confirming previous results about the existence of a MDR also
at $z \sim 0.6$ (see Sect.~\ref{s:intro}). However, the ELGs do not
avoid the central cluster regions as in low-$z$ clusters.

The evidence for segregation in velocity space is not so strong. In
fact the \sv-profiles of nELGs and ELGs are not very different (see
the right-hand panels of Fig.~\ref{f:vprofs}), in contrast with the
situation seen in low-$z$ clusters.  We jointly compare these
\sv-profiles with those predicted by NFW $M(r_n)$-models, and either
M{\L} or OM $\beta(r_n)$ models, via the $\chi^2$ method described in
Sect.~\ref{s:method}.  The best-fit solution is obtained for the M{\L}
$\beta(r_n)$ model. The best-fit value of the NFW concentration
parameter is $c=3.2_{(-1.0) -2.0}^{(+1.2) +4.6}$ (90\% c.l., 68\% c.l.
in brackets).  The $\chi^2$ vs. $c$ solution is displayed in
Fig.~\ref{f:chi2c}, bottom panel. As expected from the smaller size of
the high-$z$ sample, the solution is less well constrained than for
the low-$z$ sample.

We then adopt the $c=3.2$ NFW best-fit solution as the reference
$M(r_n)$ for the stacked high-$z$ cluster, and look for the best-fit
$\beta(r_n)$ M{\L}-model solutions, separately for nELGs and ELGs.  We
find that the best-fit anisotropy-radius parameter is identical for
the two galaxy classes, $a=0.01$, at the lower limit of the $a$-range
considered in the $\chi^2$ minimization analysis.  Unfortunately, the
solutions are poorly constrained. The 90\% upper limit for the nELGs
is $a<0.9$, that for the ELGs is at the upper limit of the $a$-range
considered, $a \leq 10.0$. The best-fit model \sv-profiles are shown
as solid lines overlaid on the observed, binned \sv$(r_n)$ in
Fig.~\ref{f:vprofs} (right-hand panels).

The solutions for the velocity-anisotropy profiles,
  $\sigma_r/\sigma_t$, are displayed in Fig.~\ref{f:beta} (right-hand
panels) and are identical for the nELGs and the ELGs. The upper limits
on the $a$-parameters translate in lower-envelopes to the
velocity-anisotropy profiles (dashed curves in the right-hand panels
of Fig.~\ref{f:beta}). Taken at face value, both the nELGs and the
ELGs have radially anisotropic orbits.  Isotropic orbits are excluded
for high-$z$ nELGs, at least outside the center, but cannot be
excluded for high-$z$ ELGs because of the large error bars.

The poor constraints on the velocity anisotropy of high-$z$ ELGs do
not allow us to draw any conclusion about their orbital evolution with
$z$. Taken at face value the results suggest that ELGs follow radially
anisotropic orbits both in high-$z$ and in low-$z$ clusters. On the
other hand, the orbits of nELGs evolve with $z$, from almost isotropic
in low-$z$ clusters to radially anisotropic in high-$z$ clusters
(compare the left-hand and right-hand panels of Fig.~\ref{f:beta}).

\section{Summary and discussion}
\label{s:disc}
We have obtained acceptable equilibrium solutions to the nELG and ELG
projected phase-space distributions with NFW $M(r_n)$ models, and with
OM and, respectively, M{\L} $\beta(r_n)$ models, for the low-$z$ and,
respectively, the high-$z$ sample. The solutions do not always provide
excellent fits to the \sv$(R_n)$ of the nELGs and ELGs, but they are
anyway always acceptable at the $90$\% c.l. Hence, the rather limited
set of models we have considered seems to be adequate for our
data-sets. Larger data-sets would be required if a wider range of
models and/or parameters is to be considered.

\begin{figure}
\centering \resizebox{\hsize}{!}{\includegraphics{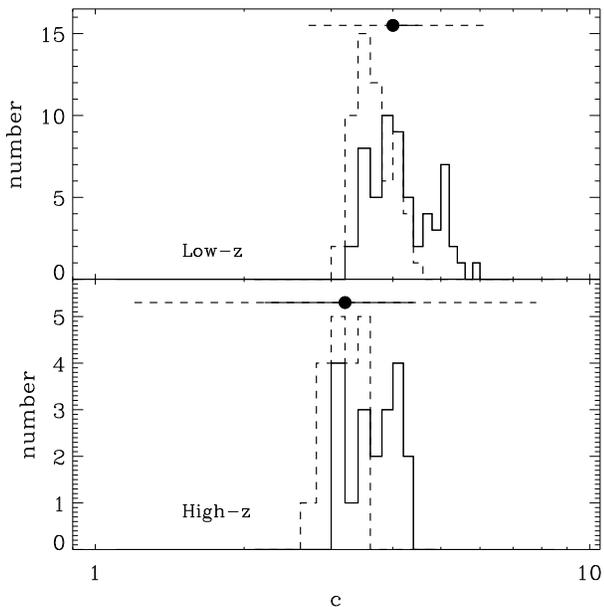}}
\caption{The distributions of predicted $c$ for the low-$z$ (upper
  panel) and high-$z$ (lower panel) cluster samples, as obtained from
  the relation of \citet{Gao+08} (solid histograms) and
  \citet{Duffy+08} (dashed histograms) applied to our cluster mass and
  redshift estimates.  The dots indicate the best-fit $c$ values from
  the Jeans analysis, and the solid and dashed lines above the
  histograms the 68\%, respectively 90\%, confidence intervals on the
  best-fit solutions.}
\label{f:c_dist}
\end{figure}

The best-fit NFW $c$ values found with the Jeans analysis are in
agreement with those expected in a $\Lambda$CDM universe for
cosmological halos with the masses and redshifts of the clusters in
our samples. In Figure~\ref{f:c_dist} we display the distribution of
the predicted $c$ values for the clusters in the low-$z$ and high-$z$
samples. The predicted $c$ values are obtained from our cluster mass
and redshift estimates, using the $c=c(M,z)$ relations of
\citet{Gao+08} and \citet{Duffy+08}.  In the same figure we also
display the best-fit $c$ value obtained with the Jeans analysis, and
the 68 and 90\% c.l. The Jeans solutions are fully consistent with the
predicted distributions of $c$ values, both at low- and at
high-$z$. 

\citet{Duffy+08} claimed that a discrepancy exists between the
predicted $c$ values for groups and clusters of galaxies and those
determined observationally using X-ray observations. Since this is not
apparent here, the claimed discrepancy may occur mostly at the group
scale \citep[see Fig.~8 in][]{Biviano08} or originate from a systematic
bias in the the X-ray-based cluster mass estimates \citep{Rasia+06}.

Our results are therefore consistent with the theoretical
$c=c(M,z)$ relation for cosmological cluster-sized halos in a
$\Lambda$CDM universe.

As far as the orbits of cluster galaxies are concerned, our analysis
suggests that these orbits become more isotropic with time. Low-$z$
cluster nELGs have nearly isotropic orbits, low-$z$ cluster ELGs have
radially anisotropic orbits outside the central cluster regions.  We
can exclude that ELGs follow isotropic orbits at the 90\% c.l. outside
$r \simeq 0.6 \, r_{200}$. 

In high-$z$ clusters nELGs and ELGs appear to be characterized by
similar orbital anisotropies, radial and almost constant with
radius. For nELGs we can exclude isotropic orbits outside $r \simeq 0.1
\, r_{200}$ at the 90\% c.l. Since low-$z$ cluster nELGs are
characterized by nearly isotropic orbits, the orbits of nELGs must
evolve from radial to isotropic with time. No significant
evolution is found for the orbits of ELGs, which remain radially
anisotropic both at high- and low-$z$.

A process capable of isotropizing galaxy orbits with time is the
secular growth of cluster mass via hierarchical accretion
\citep{Gill+04}. According to theoretical models, the growth of halo
masses occurs in two phases, an initial, fast accretion phase,
followed by a slower, smoother accretion phase \citep{LC09}. During
the fast accretion phase clusters undergo major mergers that can
induce rapid changes in the cluster gravitational potential
\citep{Manrique+03,PDdFP06,Valluri+07}, causing energy and angular
momentum mixing in the galaxy distributions, and thereby
isotropization of galaxy orbits
\citep{Henon64,LyndenBell67,KS03,Merritt05,LC09}.

Cluster-sized halos of $10^{14-15} \, M_{\odot}$ mass undergo the
transition from the fast to the slow accretion phase at $z \approx
0.4$ \citep[see Fig.6 in][]{LC09}. Hence, clusters in our high-$z$
sample are observed before the end of their fast accretion
phase. During this phase, the orbits of cluster galaxies can still
evolve, approaching isotropy. At $z<0.4$ one therefore expects to see
isotropic orbits for those galaxies that were already part of the
clusters before the end of the fast accretion phase. In our low-$z$
cluster sample we therefore expect nELGs to be characterized by
isotropic orbits, as observed. On the other hand, ELGs must be
newcomers in low-$z$ clusters, hence memory of their recent cluster
infall is still conserved in their (radially elongated) orbits.

The fact that a significant fraction \citep[$23 \pm
3$\%,][]{Poggianti+09} of the high-$z$ nELGs have a $k+a$ or $a+k$
spectral classification is supporting evidence for the fact that the
clusters of our high-$z$ sample are still in the fast accretion phase.
The spectral characteristics of these galaxies are in fact indicative
that their star formation have stopped in the $\sim 1.5$ Gyr prior to
observation \citep{Poggianti+99}. If the cessation of the
star-formation activity in these galaxies is related to their first
encounter with the cluster environment, their accretion has been
relatively recent. Hence, a large part of the cluster galaxies (and
presumably, of the cluster mass) has been assembled in the last $\sim
1.5$ Gyr.

The detected orbital evolution occurs over the same period of cosmic
time (the last $\sim 5$ Gyr) when the cluster galaxy population
undergoes a major change in its morphological mix \citep{Desai+07} and
star formation properties \citep{Poggianti+06}.  We might then be
witnessing two effects of the same underlying physical phenomenon.

The transformation of ELGs into nELGs may be at least partly
responsible for another evolutionary trend we observe in our clusters,
that of the nELG number density profile. This profile becomes less
concentrated with time, an evolution in the opposite sense to that
observed for the mass density profile.  The (projected) NFW models
fitted to the number density profile of low- and high-$z$ nELGs have
best-fit concentrations $c=2.4_{-0.2}^{+0.6}$ and
$c=7.5_{-0.9}^{+1.6}$ (90\% c.l.; see also Figure~\ref{f:nprofs}). On
the other hand, no significant evolution is found for the number
density profile of nELGs and ELGs together. This profile is dominated
by nELGs at low-$z$, but not at high-$z$. If high-$z$ ELGs transform
into nELGs with time, the nELG+ELG number density profile would not
change, but the nELG number density profile would flatten, since ELGs
are less spatially concentrated than nELGs. Also the ELG number
density profile flattens with time. This might be related to the
cluster environment growing more hostile with time, and making more
difficult for infalling field galaxies to conserve their gas as they
approach and cross the cluster centers.

The results we have obtained in the present study are based on the
still rather limited amount of available data for high-$z$ cluster
galaxies. Moreover, our high-$z$ and low-$z$ cluster samples span
quite a substantial range in masses. It will then be important to
tighten the current constraints on the orbital evolution of cluster
galaxies using future, larger spectroscopic data-sets for high-$z$
clusters, and also to re-assess such an evolution as a function of
cluster mass.  From these future analyses we will obtain a more
thorough understanding of the hierarchical assembly history and
evolution of galaxy clusters.

\begin{acknowledgements}
We acknowledge useful discussions with Giuseppina Battaglia, Alfonso
Cavaliere and Gary Mamon. We thank the anonymous referee for
  her/his useful remarks.  This research has been financially
supported from the National Institute for Astrophysics through the
PRIN-INAF scheme. This research has made use of NASA's Astrophysics
Data System.
\end{acknowledgements}

\bibliography{ediscs}

\end{document}